\documentclass[twocolumn,apjfonts]{aastex631}

\usepackage{apjfonts}

\usepackage{graphicx}
\usepackage[suffix=]{epstopdf}
\usepackage{natbib}
\usepackage{amsmath}
\usepackage{url}
\usepackage{xspace}
\usepackage{multirow}
\usepackage{booktabs}
\usepackage{color}
\usepackage[dvipsnames]{xcolor}
\usepackage{tcolorbox}
\usepackage{array, makecell}

\usepackage{hyperref}

\defcitealias{wainer_searching_2024}{W24}

\providecommand{\e}[1]{\ensuremath{\times 10^{#1}}}

\shortauthors{Wainer et al.}

\begin{document}

\title{Searching for Stellar Activity Cycles using Flares II: The TESS CVZ}

\correspondingauthor{Tobin M. Wainer}
\email{tobinw@uw.edu}

\author[0000-0001-6320-2230]{Tobin M. Wainer}
\affiliation{Department of Astronomy and the DiRAC Institute, University of Washington, 3910 15th Avenue NE, Seattle, WA 98195, USA}

\author[0009-0005-9247-0863]{Holland M. Stuart}
\affiliation{Department of Astronomy and the DiRAC Institute, University of Washington, 3910 15th Avenue NE, Seattle, WA 98195, USA}

\author[0000-0001-5455-6678]{Guadalupe Tovar Mendoza}
\affil{Department of Physics and Astronomy, Johns Hopkins University, 3400 N Charles St, Baltimore, MD 21218, USA}

\author[0000-0003-0174-0564]{Andrew R. Casey}
\affiliation{Center for Computational Astrophysics, Flatiron Institute, 162 Fifth Avenue, New York, NY 10010, USA}
\affiliation{School of Physics and Astronomy, Monash University VIC 3800, Australia}

\author[0000-0001-6147-5761]{Tom Wagg}
\affiliation{Center for Computational Astrophysics, Flatiron Institute, 162 Fifth Avenue, New York, NY 10010, USA}

\author[0000-0002-0637-835X]{James R. A. Davenport}
\affiliation{Department of Astronomy and the DiRAC Institute, University of Washington, 3910 15th Avenue NE, Seattle, WA 98195, USA}

\begin{abstract}
    Magnetic activity cycles provide a fundamental constraint on stellar dynamos, but remain difficult to identify beyond the Sun. However, recent studies have shown that flares offer a unique tracer of activity cycle behavior. In this study, we use seven years of short-cadence observations from the Transiting Exoplanet Survey Satellite (TESS) for over 14,000 stars in the Continuous Viewing Zone to search for long-term changes in flare activity. For each star, we perform injection and recovery tests and provide well-characterized completeness limits for flare detection thresholds, and flare finding results. From this search, we identify 17 stars with evidence of long-term variability in flare rate, synonymous with activity cycle behavior. These candidates span a range of effective temperatures, rotation periods, and flare-variability morphologies. One G-type star, TIC 167344043, stands out as the clearest solar-like case, despite rapid rotation and superflare activity. 
    Our results identify candidate activity cycles in stars that are more rapidly rotating and flare-active than the typical stellar-activity-cycle targets in the literature. These systems probe a new regime where stellar dynamos are still evolving, providing critical constraints on when cycle-like magnetic variability first emerges.
\end{abstract}



\section{Introduction}



The Sun's 22-year magnetic activity cycle remains one of the most critical constraints on dynamo theory \citep[e.g.,][]{babcock_topology_1961}. This cycle modulates the Sun’s surface magnetism and manifests across a range of observables—including sunspot coverage, chromospheric line emission, flare rates, and total solar irradiance—each varying over months to decades \citep{charbonneau_solar_2014}. While solar activity has been tracked in detail for centuries \citep{eddy_book_1980, usoskin_history_2017}, detecting analogous cycles in other stars remains challenging. Most cycle detections rely on chromospheric diagnostics such as CaII H\&K emission \citep[e.g.,][]{baliunas_chromospheric_1995, hall_activity_2007, baum_five_2022, isaacson_california_2024}, which require multi-decade spectroscopic monitoring. These surveys, though powerful, are typically limited to bright, nearby stars and small samples.

In contrast, photometric flare monitoring offers a relatively new approach to activity cycle detection \citep{scoggins_using_2019, feinstein_evolution_2024,wainer_searching_2024}. Flares arise from magnetic reconnection in stellar atmospheres \citep{benz_physical_2010, kowalski_time-resolved_2013}, and their occurrence rates correlate with both spot coverage and large-scale field strength \citep{davenport_kepler_2016, feinstein_testing_2022}, which depend on the phase of the solar cycle. On the Sun, flare rates vary in lockstep with the activity cycle \citep[e.g.,][]{veronig_temporal_2002, wainer_searching_2024}, suggesting that similar trends might be detectable in other stars. With the long baselines available from Kepler \citep{borucki_kepler_2010} and TESS \citep{ricker_transiting_2015}, photometric flare surveys offer a promising avenue for constraining magnetic variability across thousands of stars. 

However, cross-telescope comparisons of stellar flare rates are notoriously difficult. While both Kepler and TESS provide high-precision photometry suitable for flare detection, the differences in instrument design—including cadence, aperture, and bandpass—can significantly impact flare recovery and energy estimates \citep[e.g.,][]{kowalski_time-resolved_2013, howard_evryflare_2020, raetz_long-term_2024}. Even with consistent flare identification pipelines, subtle differences in instrumental response, and bandpass, can lead to systematic offsets in flare frequency distributions (FFDs) and their fitted slopes. 

Luckily, TESS data alone is now beginning to achieve a long enough baseline to adequately measure long-term flare variability. In particular, stars in the continuous viewing zones (CVZ) are now observed across more than a dozen Sectors over five or more years. This sustained coverage enables population-level flare studies that are both statistically powerful and internally consistent, avoiding the cross-calibration challenges that arise when combining multiple telescopes. 

Recent work by \citet{feinstein_evolution_2024} demonstrates the potential for detecting long-term variability in stellar flare rates using TESS data alone. Their sample of eleven young stars, each observed in multiple TESS Sectors over a five-year baseline, showed year-to-year changes in flare activity, an encouraging step toward detecting stellar magnetic cycles from flare monitoring. Our work builds directly on these results by focusing on stars with far denser temporal coverage, and specific treatment of flare variability metrics. 

In \citet[][hereafter W24]{wainer_searching_2024}, we introduced a statistical framework for quantifying variability in stellar flare activity over time. Rather than relying solely on absolute flare rates, we focused on measuring changes in the FFD, incorporating the effects of stochastic sampling, incompleteness, and limited event counts. We demonstrated how flare rate fluctuations could arise from finite sampling even in the absence of true magnetic variability, and presented a method for assessing the significance of apparent changes as a function of observing baseline. This approach enabled a consistent way to evaluate flare variability both within and across light curves, providing a foundation for identifying candidate activity cycles in time-resolved flare data. While that study focused on a single target, the methods developed in \citetalias{wainer_searching_2024} are directly extensible to larger samples, such as those with dense temporal coverage, and long baselines.

In this study, we apply this statistical framework developed in \citetalias{wainer_searching_2024} to a sample of over 14,000 stars in the TESS CVZ. These stars have been observed across a minimum five-year baseline of the TESS mission. We identify and characterize flares across this entire sample, constructing time-resolved FFDs and assessing their evolution using a consistent set of statistical metrics. By combining long-duration coverage with a homogeneous observing strategy, this dataset provides an ideal laboratory for searching for stellar activity cycles through flare variability.

This paper is structured as follows. We first discuss our sample of stars in Section~\ref{sec:tess_data}, while introducing our flare finding methodology in Section~\ref{sec:flare_finding}, and then how we characterize variability in these flare stars over time in Sections~\ref{sec:flarechar} and \ref{sec:var_finding}. We present our results in Section~\ref{sec:results}, and discuss our sample candidates is Section~\ref{sec:best_discussion}. We then compare to previous studies and discuss caveats in Section~\ref{sec:discussion_section}.

\section{TESS Data}\label{sec:tess_data}

The Transiting Exoplanet Survey Satellite (TESS) collects near-continuous photometric observations in $\sim$27-day intervals known as Sectors, with 13 Sectors typically observed each year \citep{ricker_transiting_2015}. During its primary mission (Cycles~1 and~2), TESS monitored nearly 200{,}000 stars selected from the TESS Input Catalog \citep[TIC;][]{stassun_tess_2018} at a 2-minute cadence. These observations were obtained through a broad-band filter covering wavelengths from approximately 600 to 1000~nm, centered near the Cousins $I$-band.

In the extended mission (Cycles~3--8), TESS has continued this observing strategy while expanding its temporal coverage across the sky. Stars located near the ecliptic poles fall within the mission's continuous viewing zones (CVZs), where they can be observed in nearly every Sector of a given hemisphere. As a result, many targets in the CVZ have accumulated long, nearly uninterrupted baselines spanning over 7 years to date. This sustained coverage makes the CVZ regions especially well suited for studying long-term stellar variability, including potential magnetic activity cycles traced by flare rates.

\subsection{Stellar CVZ Sample}

\begin{figure}
    \centering
    \includegraphics[width=0.47\textwidth]{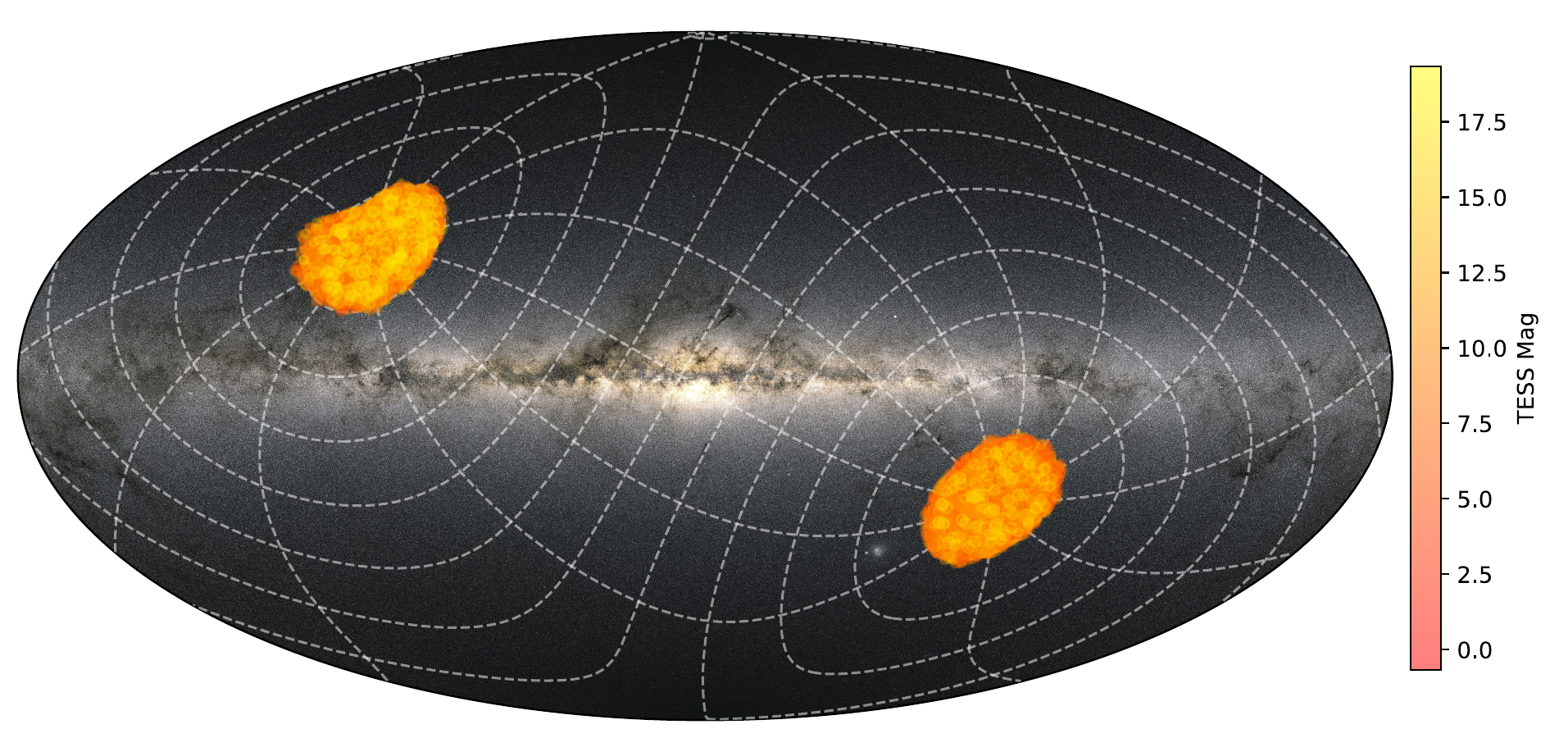}
    \caption{14,385 CVZ targets that are analyzed in this work. The grid represents RA and DEC, showing where each of our targets are in the sky. Each point is colored by its TESS magnitude. This figure was made with \texttt{mw-plot}, with the background image from Gaia's all-sky optical view; credit: ESA/Gaia/DPAC, CC BY-SA 3.0 IGO \citep{gaia_collaboration_gaia_2018}. }
    \label{fig:scatter}
\end{figure}

We gathered all targets from the first 5 years of TESS 2-minute observations (Sectors 1--69), which included 483,231 unique TIC objects across the full sky. From these, we selected stars with at least 13 unique Sectors of data, which naturally limited our sample to the CVZ. This identified a sample of 14,385 stars, which is dominated by nearby field dwarfs. These stars have between 13 and 36 Sectors of TESS 2-minute data, with a median of 22 Sectors of data up through Cycle 5.
The location on sky of our sample is visualized in Figure~\ref{fig:scatter}.

\section{Methods} \label{sec:methods}

Our analysis on the long term variability of stellar flares is predicated on our ability to accurately find and characterize individual flare events. In this work, we build on the methods from \citetalias{wainer_searching_2024}, which demonstrated how TESS flare samples can be characterized quantitatively by combining uniform flare detection, injection–recovery tests, and completeness-corrected FFDs. For a detailed description of these methods, we refer the reader to \citetalias{wainer_searching_2024}. Here, we provide a summary, and describe deviations from \citetalias{wainer_searching_2024} adopted  the current study of 14000 stars compared to the single star case. 

\subsection{Flare Identification} \label{sec:flare_finding}

We identify flares using the convolutional neural network package \texttt{stella} \citep{feinstein_stella_2020}, following the methodology developed in \citetalias{wainer_searching_2024}. For each TESS light curve, we computed the average probability from ten trained \texttt{stella} models \citep[as in][]{feinstein_evolution_2024}. We identified flare candidates as regions with three or more consecutive time steps having probabilities $>30\%$, and merged events within 40 minutes of one another to account for complex morphology \citep{zhu_complex_2015}. This thresholding and merging procedure was applied uniformly across all stars and Sectors in the sample.

\begin{figure*}
    \centering
    \includegraphics[width=0.975\textwidth]{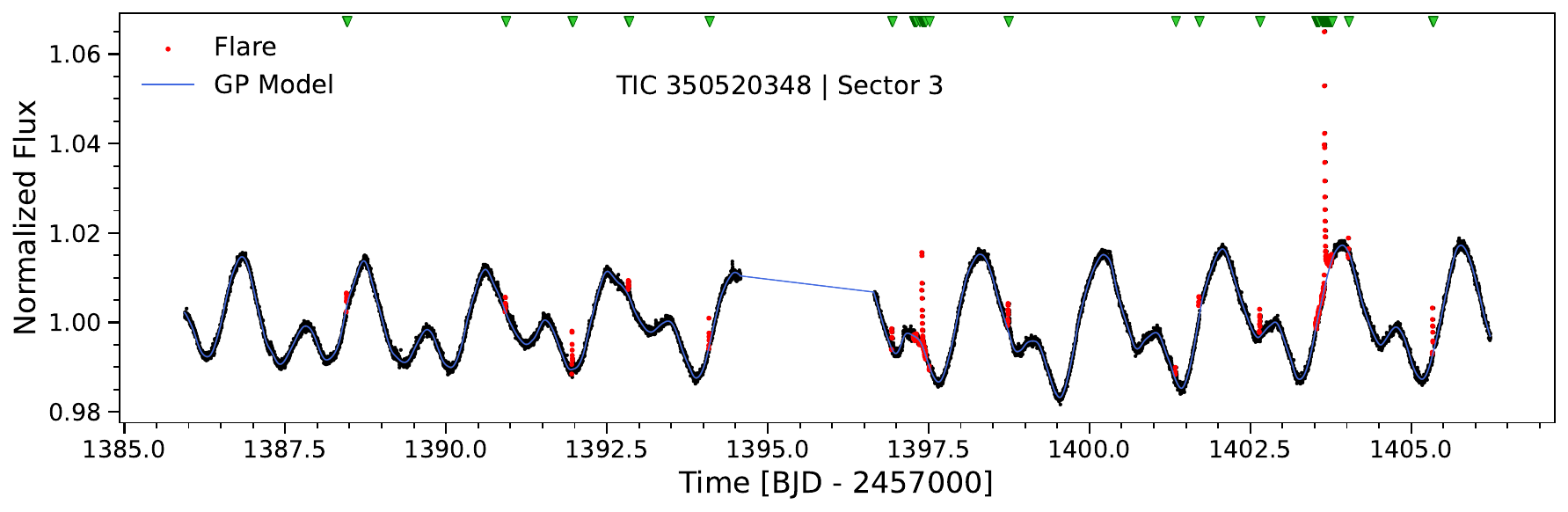}
    \caption{Light curve for TIC350520348 with identified flares. The y-axis is normalized by the median flux of the light curve. Points that are considered to be flares by \texttt{stella} are shown in red, and there are green markers at the top of the plot for each identified flare. Shown in blue is the GP fit. }
    \label{fig:lc_example}
\end{figure*}

We show an example of our flare-finding results for a single star and Sector in Figure~\ref{fig:lc_example}. The black points show the normalized TESS light curve, and the red points indicate points identified as part of flare events. The green markers at the top of the panel mark the times of the detected flares, and the blue curve shows the Gaussian process model for the underlying stellar variability (Section~\ref{sec:flarechar}). This example illustrates how \texttt{stella} is able to detect flare brightening from the broader variability of the light curve for a high signal-to-noise star.

For a study analyzing flares of over 14,000 stars, the benefits of automated flare finding cannot be understated. Looking through over 100,000 sector light curves manually to try and identify flares is far too cumbersome a task. However, simply applying a machine learning algorithm uniformly to a large data-set of stars each with unique rotation periods and star spot modulation, luminosities, and photometric uncertainties is non-trivial, and there are several instances where \texttt{stella} fails. 

Specifically, we find several cases in which the events identified by \texttt{stella} are clear false positives rather than bona fide stellar flares. These failures modes are not arbitrary, but occur for stellar light curves that have long been problematic for automated flare searches \citep[e.g.][]{davenport2019}. Periodic and pulsational variables, including eclipsing binaries and RR Lyrae stars, are known sources of false flare detections in both classical and machine-learning-based pipelines \citep{vida_finding_2021,aschwanden_self-organized_2021,jia_fcn4flare_2025}. In eclipsing binaries, sharp changes during eclipse ingress or egress can mimic the impulsive morphology of a flare, while in RR Lyrae and other pulsators, rapid changes near maximum light can produce similarly flare-like structure.

We additionally identify false positives in stars with very short rotation periods and high-amplitude starspot modulation, where the ``cuspy'' rotational signal is sometimes misclassified as a flare. False positives also become more common when the flux uncertainty is of the same order as the intrinsic variability amplitude of the star, since low-level stochastic structure can then be mistaken for impulsive brightenings (i.e. flares). We discuss the handling of such false positives more in Section~\ref{sec:results}, but emphasize that flare finding is not a solved problem, and more work is needed if the community wants to progress in the era of big data.  

\subsubsection{Completeness Quantification Through Injection and Recovery}\label{sec:completeness}

An important component of flare distribution studies is an accurate accounting of the flares recovery completeness \citep{davenport_kepler_2016}. Here, we evaluate flare detection completeness using injection-recovery tests. Synthetic flares were generated using the analytic flare model from \citet{tovar_mendoza_llamaradas_2022}, with amplitudes drawn logarithmically between the photometric noise floor and 50 times that value. For each TESS Sector, we performed $N=30$ injections per trial across $R=25$ repeats, inserting flares at random non-flaring times and rerunning the full \texttt{stella} pipeline. Recovery was defined by overlap with the injected time window. This process typically resulted in $>10^4$ injection-recovery tests per star.

We computed completeness as a function of equivalent duration, $ED$, and fit a logistic model:
\begin{equation}\label{eq:completeness}
\text{completeness}(x) = \frac{0.96}{1 + \exp[-k(\log_{10}(ED) - x_0)]},
\end{equation}
where $k$ and $x_0$ were fit per Sector. The 50\% completeness threshold is then adopted as the lower limit for flare inclusion in the our analysis. 

As discussed in \citetalias{wainer_searching_2024} and \citet{feinstein_flare_2020}, \texttt{stella} never quite reaches 100\% completeness. This is primarily because \texttt{stella} will not search for flares within 300 time steps of gaps in the data, which are present within every TESS light curve. This effect can be seen in Figure~\ref{fig:lc_example}, where at $\sim1394.5$ there is a small flare that is not detected due to its proximity to the characteristic TESS mid-Sector down-link gap. We therefore follow \citetalias{wainer_searching_2024} and set the maximum value of the the logistic completeness threshold to 0.96 to reflect these affects. 

\subsection{Flare Characterization}
\label{sec:flarechar}

We evaluate each star's flare activity using four complementary diagnostics: (1) the total number of detected flares per Sector, (2) the FFD, (3) the intercept $\beta$ of the power-law fit to the FFD, and (4) the integrated flare luminosity ratio $L_{\mathrm{fl}}/L_{\mathrm{bol}}$ \citep[e.g.,][]{feinstein_evolution_2024}. 

We emphasize that any single one of these metrics only provides a partial picture of the overall flare behavior of the star. We therefore use all metrics equally in our analysis looking for tracers of long-term magnetic variability.

All flare energies in this work are measured in terms of equivalent duration (ED), defined as the time integral of the relative flux enhancement above quiescence \citep{gershberg_results_1972,hunt-walker_most_2012} in units of seconds. Opposed to using flare energies in units of erg per second, this approach eliminates the use of uncertain stellar luminosities, and ensures that our analysis remains internally consistent across the full sample.

To accurately measure a flare's ED, we first subtract a baseline model from the light curve, which accounts for underlying stellar variability, such as starspot modulation. This baseline model is generated using a two-component Gaussian Process (GP) model implemented with \texttt{tinygp} \citep{foreman-mackey_dfmtinygp_2024}. We mask all flare events identified by \texttt{stella} and fit the remaining light curve using a rotation-inspired kernel with both short- and long-timescale cosine components. The GP is trained on the masked data via likelihood optimization using \texttt{jaxopt}, \citep{jax2018github} with iterative sigma clipping to exclude outliers. Once trained, we evaluate the GP across the full light curve, through masked flares, to estimate the quiescent flux at all time steps. This model provides a fast, smooth, and physically motivated baseline for computing EDs, even in stars with strong or highly variable rotational variability.

As is standard, we construct the FFD for each TESS Sector as the reverse cumulative distribution of flare EDs, corrected for completeness via our injection-recovery tests (Section~\ref{sec:tess_data}). Following \citetalias{wainer_searching_2024}, we adopt the 50\% completeness threshold (calculated with Eq.~\ref{eq:completeness}) as the minimum ED for inclusion in the FFD, since lower completeness levels introduce significant biases in the inferred flare rates and slopes from the completeness correction. 

The FFDs are modeled as power laws, which in log-space takes the form:
\begin{equation}
y = \alpha x + \beta,
\end{equation}
where $y$ is the log cumulative flare rate (in flares per day), x is the log10 ED (in seconds), $\alpha$ is the power-law slope, and $\beta$ is the intercept.

For each star, we determine a global best-fit slope $\alpha$ using all available flares across all Sectors, then fix $\alpha$ and fit for $\beta$ to the completeness corrected data in each individual Sector. This decision follows directly from \citetalias{wainer_searching_2024}, and is motivated by evidence to suggest the slope of the power law can be dependent on the type of star \citep[e.g.,][]{feinstein_testing_2022}, but consistent for a given star. The fixed-slope assumption may be imperfect, particularly if the underlying flare energy distribution changes as the flare rate evolves. However, allowing the FFD slope to vary independently in each sector can produce systematically unstable, and nonphysical fits when the flare sample is small and stochastic sampling dominates. We choose to adopt this fixed slope strategy for consistency, and to ensure that changes in $\beta$ reflect genuine evolution in flare activity, rather than noise in fitting the FFD slope. 

While $\beta$ encodes information about the full flare distribution, we also examine the total flare counts per Sector and the integrated flare power, $L_{\mathrm{fl}}/L_{\mathrm{bol}}$, for the non-completeness corrected flare distribution. This combination of metrics provides a useful cross-check, especially for stars with fewer flares, and allows for comparison with prior studies of young and active stars. 

\subsubsection{Uncertainties on Metrics}
Although the flare activity metrics discussed above have each been used throughout the literature as indicators of stellar activity, their interpretation depends critically on a corresponding uncertainty model. The flare activity metrics used throughout this work are inferred from finite realizations of an intrinsically stochastic process, and the measured values depend not only on the underlying flare distribution, but also on the observing baseline, the number of detected events, telescope systematics, photometric uncertainty, and the energy-dependent completeness of the flare search. Quantifying the uncertainty in each derived metric is therefore an essential part of the analysis.

While the literature increasingly reports uncertainties on flare quantities such as flare rates, FFD slopes, and completeness functions \citep{howard_flaring_2022,gao_correcting_2022}, uncertainties are often not propagated through to the final time-resolved population metrics. In the search for activity cycles using flares, far less attention has been given to quantifying the uncertainty on the final sector- or epoch-level flare rate metrics themselves. In this work, we treat uncertainty estimation as a central part of the analysis, and argue that this uncertainty characterization should become standard practice in time-domain flare studies. Without explicit uncertainties on the final reported metrics it is difficult to determine whether an apparent change in activity is astrophysical, or simply the level of scatter expected from finite sampling, photometric uncertainty, or completeness effects. 

For the three metrics we derive and analyze in our search for long term variation in flare rate ($\beta$, $F_{\rm flare}/F_{\rm bol}$, and N Flares), we describe the derived uncertainties below. 

\textit{$\beta$:} The uncertainty on $\beta$ is taken from a combination of the derived fit uncertainty, and the uncertainty from observing for a finite period of time. First, we consider the covariance matrix returned by the \texttt{curve\_fit} function in \texttt{scipy} \citep{virtanen_scipy_2020}), the fitting procedure used to derive FFD parameters throughout the analysis. In our case, the fit incorporates the Poisson counting uncertainties on the cumulative FFD values, so the formal uncertainty on $\beta$ reflects the statistical uncertainty associated with the finite number of flares contributing to that interval. Because each individual TESS sector spans only $\sim27$ days, we then add an additional uncertainty directly to the sector-level $\beta$ uncertainty to account for the extra variance introduced by the short baseline. The exact value of this uncertainty is defined by the length of the baseline considered, derived in \citetalias{wainer_searching_2024}. For a single sector, this equals 0.25, while for sector groups of 3 it is 0.14. 

\textit{$F_{\rm flare}/F_{\rm bol}$:} The uncertainty on the integrated flare power is derived directly from the photometric flux errors. For each flare, we take the flux error through the flare and divide by the median stellar flux to place the uncertainty on the same relative normalization used for the ED calculation. We then compute an uncertainty-ED using the same trapezoidal-sum procedure used to calculate the flare ED itself, and propagate that through the sector duration normalization used to derive $F_{\rm flare}/F_{\rm bol}$. The reported uncertainty then reflects the photometric uncertainty associated with the integrated flare signal rather than only the number of detected events. This photometric uncertainty is especially important for integrated flare power, which is sensitive to the cumulative contribution of flux across each flare. 

\textit{flare counts:} The uncertainty on the flare count in each sector is taken to be the square root of the number of detected flares, corresponding to the usual Poisson counting uncertainty appropriate for discrete events. This provides the natural statistical uncertainty on the total number of flares contributing to a given sector or epoch. 

\subsubsection{To Bin or Not to Bin?} \label{sec:binning}

A central choice in flare studies is how aggressively to bin the flare measurements in time. Combining more consecutive sectors into a single data point reduces the effects of stochastic flare sampling, and correspondingly decreases the uncertainty on the derived activity metrics \citepalias[][]{wainer_searching_2024}. This is the logic behind the coarse time-binning adopted in some previous work, where years of data are combined into a single measurement \citep{feinstein_evolution_2024}. For studies focused on multi-year variability, this approach can be useful, since it suppresses short-timescale scatter and emphasizes longer-term trends.

The drawback is such coarse binning can also smooth over real changes in flare activity on much shorter timescales. If the flare rate varies substantially from month-to-month, combining many sectors together will wash out structure and make brief, but astrophysically meaningful, enhancements/declines in activity difficult to identify. 

In this work, we therefore consider both perspectives. We analyze the individual sector measurements directly, which retain the highest time resolution, but also the largest sensitivity to stochastic flare sampling. We also construct binned measurements by combining any sectors separated by less than 150 days into a single data point, allowing us to trace broader changes in activity over longer baselines. Finally, we adopt an intermediate approach in which sectors are grouped into sets of approximately three consecutive sectors, with some groups containing two or four sectors when necessary. This hybrid scheme is intended to capture the advantages of both methods, reducing the stochastic scatter associated with single-sector measurements while still preserving much of the shorter-timescale variability that is lost when the data are binned too coarsely. Each of these binned data are considered equally, while particularly convincing evidence is prominent in all three regimes.  

\subsection{Variability Finding} \label{sec:var_finding}
Although our CVZ sample contains 14,385 stars, only a subset has the flare activity, temporal coverage, and data quality needed for a meaningful flare variability analysis. Recently, \citet{stelzer_flares_2022} found detectable optical flares in only about 32\% of their TESS-observed M-dwarf sample, illustrating that even among magnetically active low-mass stars, flaring stars suitable for this kind of analysis represent only a subset of the full population.

We therefore carried out the search for stellar activity cycles in a multi-stage pipeline, which we describe in Table~\ref{tab:cuts}, and explain in more detail below. The first stage was applied to the full sample of 14,385 stars observed across TESS Cycles 1–5. This step included data download, flare identification with \texttt{stella}, and injection-recovery completeness testing, which established the basic flare properties for every target. The second stage was more computationally intensive, requiring GP fitting, completeness modeling, FFD construction, and the derivation of the flare activity metrics. We therefore only run this second stage for a subset of stars that passed an initial set of baseline cuts.

Specifically, after the flare finding stage was completed for all 14,385 stars, we perform a series of cuts designed to remove stars with too few flare detections or too little time coverage to support a meaningful search for long-term variability. First, we required a star to have more than five detected flares in more than five sectors, and that its light curve span a baseline longer than 1000 days. This initial selection reduced the sample to 4,119 stars, which were then passed to the second stage of the pipeline.

Following the second-stage analysis, we applied a stricter set of cuts based on the completeness-corrected flare sample. We required that each star have at least three flares above the 50\% completeness limit in at least five sectors, and that the total baseline exceed 1200 days. We also removed any star ``confirmed'' to be an EB in the TESS EB catalog \citep{prsa_tess_2022}, 
and a set of stars for which \texttt{stella} identified more than 100 flares per sector, a common failure mode for the specific failure modes discussed in Section~\ref{sec:flare_finding}. These cuts yielded a sample of 1,775 stars, which we then subjected to the visual inspection procedure described below.

\begin{table}
    \centering
    \begin{tabular}{l c l}
        \hline
        Sample & $N_{\rm star}$ & Conditions \\
        \hline
        \multirow{2}{*}{CVZ sample} & \multirow{2}{*}{14,385} & Contained in TESS CVZ \\
                                        &                         & Observed in cycles 1--5 \\
        \addlinespace[1em]
        \multirow{2}{*}{Stage 2 sample} & \multirow{2}{*}{4,119}  & $\ge 5$ flares in $\ge 5$ sectors \\
                                        &                         & Light curve baseline $\ge 1000$ days \\
        \addlinespace[1em]
        \multirow{5}{*}{\shortstack[l]{Visual inspection\\sample}}   & \multirow{5}{*}{1,775}  & $\ge 3$ flares above 50\% \\
                                        &                         & completeness limit in $\ge 5$ sectors \\
                                        &                         & Light curve baseline $\ge 1200$ days \\
                                        &                         & Not confirmed EB in \citet{prsa_tess_2022} \\
                                        &                         & $\le 100$ flares per sector \\
        \addlinespace[1em]
        \multirow{1}{*}{Vetted sample}    & 75                      & Mean and median visual score $\ge 2$ \\
      \hline
    \end{tabular}
    \caption{A summary of the conditions required for a given sample of TESS stars. $N_{\rm star}$ is the number of stars that meet the conditions. The conditions are cumulative, such that each sample meets all of the conditions of the previous sample.}
    \label{tab:cuts}
\end{table}

\subsubsection{Visual Inspection Of Flare Activity Metrics in a Search for Long Term Variability}

\begin{figure*}[ht]
    \centering
    \includegraphics[width=0.95\textwidth]{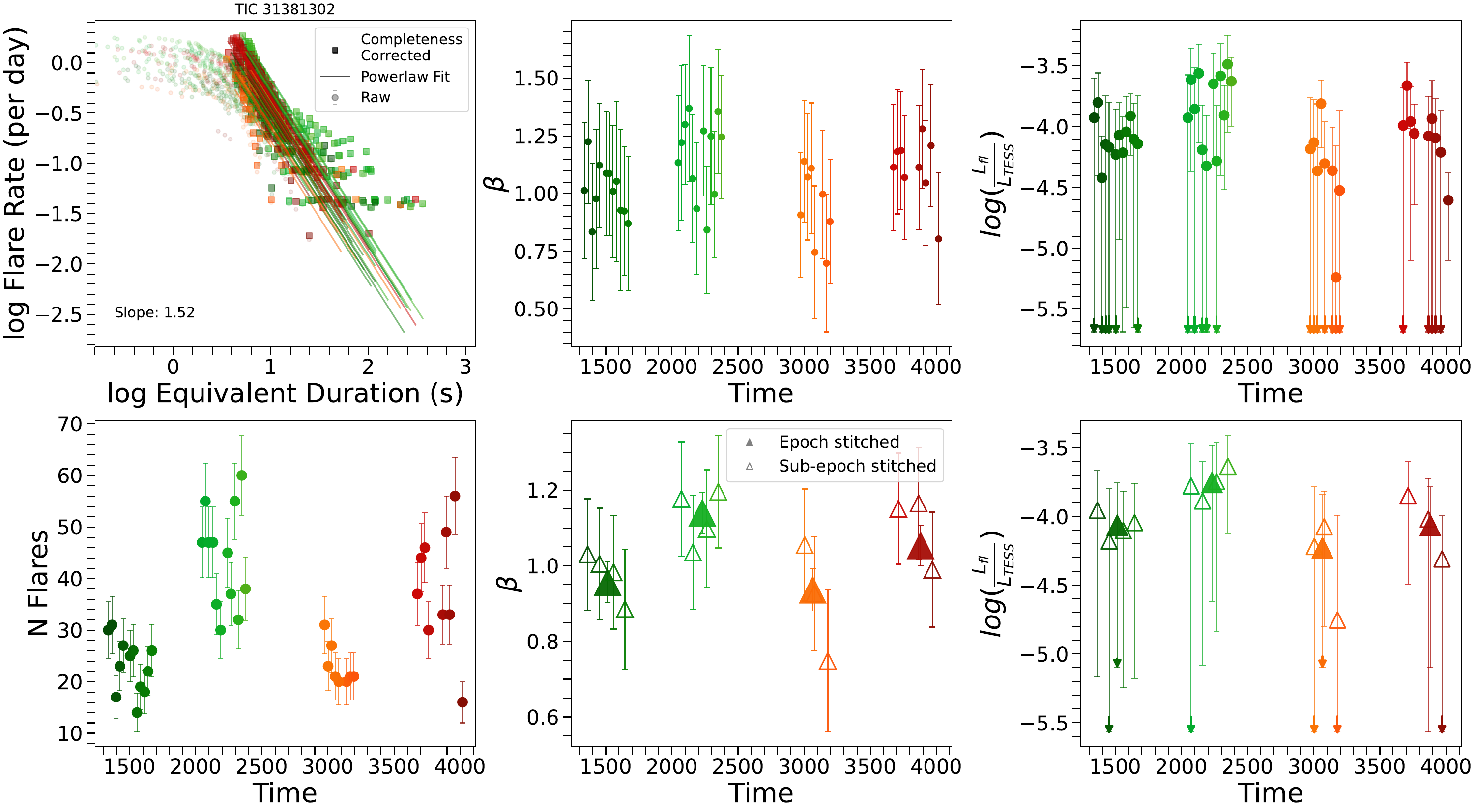}
    \caption{Example six-panel diagnostic plot used for visual inspection, shown here for TIC 31381302. Color encodes time, progressing from green to orange to red. The upper-left panel shows the FFD, and fixed-slope power-law fits, whose intercepts define $\beta$. The upper-middle and upper-right panels show the sector-by-sector evolution of $\beta$ and $L_{\rm fl}/L_{\rm TESS}$, respectively, while the lower-left panel shows the raw number of detected flares per sector. The lower-middle and lower-right panels show the same $\beta$ and $L_{\rm fl}/L_{\rm TESS}$ measurements after stitching sectors into larger time bins. Downward arrows in the $L_{\rm fl}/L_{\rm TESS}$ panels indicate measurements whose lower uncertainty bound extends below zero in linear space, marking cases where the inferred flare power is not significant relative to the photometric uncertainties. TIC 31381302 is classified as a ``Candidate'' as discussed further in Section~\ref{sec:results}.}
    \label{fig:good_diagnostic_plot}
\end{figure*}

For each star, we evaluate the flare metrics in the form of a single, 6-panel diagnostic plot. This plot shows: (1) the completeness-corrected FFD and per sector power-law fits, (2) evolution of $\beta$ over time, (3) the total flare energy per unit quiescent luminosity, and (4) the flare count per Sector. We computed these metrics for individual sectors, as well as sectors that are combined into groups (Section~\ref{sec:binning}). An example of the resulting 6-panel plot forming the base of our visual inspection is shown in Figure~\ref{fig:good_diagnostic_plot}.

In all panels, color encodes time which moves from green to orange to red, allowing measurements to be tracked consistently across the panels \footnote{We note that the color-map used here was optimized for the co-authors which performed the visual inspection, which included a specific color deficiency. We choose to show the 6-panel plots as they were for the visual inspection, while acknowledging that some red-green color blind readers may struggle. As a part of the data-behind the figure, which we present in Appendix~\ref{appendix_1}, we include code to take the data file and make these 6-panel plots, where we add an option to change the color-map if readers are so inclined.}. We show both the individual-sector measurements and the stitched measurements, where sectors are combined into larger groups (Section~\ref{sec:binning}). The stitched points are plotted with distinct marker styles to separate them from the single-sector values, and are intended to reduce the effects of stochastic flare sampling while preserving longer-term variability.

The upper-left panel shows the FFDs themselves. For each Sector, the raw FFD is shown with faint points, while the completeness-corrected flare sample is shown with larger square markers. The solid lines depict the best fit, fixed-slope power law to the corrected FFD in each Sector (Section~\ref{sec:flarechar}). 
The intercept of this fit defines $\beta$, which is shown in the middle panels. The upper-middle panel shows the evolution of $\beta$ with time for the individual sectors, while the lower-middle panel shows the same quantity for the stitched groups. The lower-left panel shows the flare count per sector. 

The right panels then show the evolution of the integrated flare power, $L_{\rm fl}/L_{\rm TESS}$, again with the individual-sector on top and and stitched measurements on bottom. A particularly useful feature of the $L_{\rm fl}/L_{\rm TESS}$ panels is the use of downward arrows for measurements whose lower uncertainty bound would extend below zero in linear space. These cases indicate that the measured flare power is not significant relative to the underlying photometric uncertainties, even if a formal positive value can still be computed. In practice, these arrows provide a quick visual indicator that the inferred integrated flare output in that sector or stitched group are consistent with noise. 

Visual inspection of these figures then allow us to assess whether apparent changes in flare activity are reflected consistently across all metrics. In practice, the combination of individual-sector and stitched measurements is especially useful during visual inspection. The individual sectors preserve the highest time resolution and are most sensitive to short-timescale changes in activity, while the stitched groups suppress some of the stochastic scatter expected from finite flare sampling. Comparing the two therefore helps distinguish isolated sector-to-sector fluctuations from broader, coherent changes in flare behavior.

For the 1775 stars that were analyzed through visual inspection, three Co-authors independently ranked their interpretation of these figure based on how strongly each star exhibited long term trends in flare activity. This search was done using the Zooniverse website \citep[e.g.,][]{trouille_zooniverse_use_2020}, which randomized the order for each person, a feature that proved useful since long classifying stretches could lead to potential bias in a single team member over time. 
Classifiers had the option of assigning a score from -1 to 3, corresponding to the strength of signal of long term variability in flare activity across all panels. 3 was set to denote a ``definite'', while 0 was then ``definitely not``, with 1--2  corresponding to different strengths of ``maybe''. A score of -1 indicated potential data quality issues, such as order of hundreds of spurious flares in a single sector, or an issue with the GP fitting as denoted by the reduced $\chi^2$ residual of the GP and data being greater than 8, which produced an error message printed on the figure. 

Once all stars were classified by the three team members, we calculated both the mean and median score for each. These two summary statistics capture complementary information about the level of agreement among classifiers. The mean reflects the overall strength of the evidence across all three classifications, while the median provides a more robust measure of the consensus.

\subsubsection{Independent Verification of Top Candidates and Including all Available Data} \label{sec:cycle7_add_on_and_vis}

After the visual inspection described above, we performed one final verification step for the most promising systems. For any star that appeared moderately interesting in the Cycle 1–5 analysis, which we considered to be having both a mean and median visual score greater than 2, 
we extended our analysis to include all available data through TESS Cycle 7. This additional processing was not feasible for the full CVZ sample, or even for the full visual inspection sample of 1775, due to the high computation cost. However, even though all stars could not be analyzed with such scrutiny, the newer TESS cycles provide a clear benefit to the study. With this additional data, the final assessment of the top candidates is as complete as possible using currently available data. In several cases, an additional epoch can distinguish between an apparent short-baseline fluctuation and a more coherent change in flare activity. Conversely, the new sectors can also weaken an initially promising case if the apparent trend disappears once a longer baseline is included.

We therefore define a ``Vetted sample'' for which we obtain data through Cycle 7. For this Vetted sample of 75 stars, we then repeated the visual inspection from Section~\ref{sec:var_finding}, and scoring procedure using the updated Cycle 1–7 diagnostic plots. As part of this final review, we also inspected the light curves directly to identify spurious flare detections, poor GP fits, or other data-quality issues that could mimic changes in the flare metrics. 

Our final rankings for the Vetted sample fall into four categories, which we deem ``Best'', ``Candidate'', ``Ambiguous'', and ``Sneaky Trash''. The ``Best'' group are our top candidates for long term variability in flare rate, which show clear, coherent, and patterned trends in all flare metrics. The second ``Candidate'' group are stars where there is clear variation, but an obvious pattern to the variability doesn't immediately pop out.
For the third ``Ambiguous'' group, while there are clear and obvious flares, it is unclear whether there is any robust variability in the flare metrics. Finally, our ``Sneaky Trash'' stars surface potential issues with the flares that \texttt{stella} identified, most notably false positive flare identifications from rapidly rotating stars or unconfirmed eclipsing binaries.  

\section{Results}
\label{sec:results}

In this section, we present the results of our flare search. We present results for each of the steps described in Section~\ref{sec:methods}, in order for replication of our detailed, human-based flare rate activity search. 

We begin with the full automated flare-search output for the TESS CVZ sample. These data define the observational foundation for the rest of the analysis, which we present in Table~\ref{tab:full_sample}. For each star, we list the number of TESS Sectors observed across Cycles 1–5, the total time baseline covered by those observations, the total number of flares identified by \texttt{stella}, the number of sectors with at least five detected flares, and the 50\% completeness limit from the injection-recovery tests (Section~\ref{sec:flare_finding}). Together, these quantities describe both the available time coverage and the effective flare-detection sensitivity for each target.

We emphasize that Table~\ref{tab:full_sample} reflects the automated flare search products rather than a vetted flare catalog. At this stage, individual flares have not been inspected by eye, and false positives are expected. These data therefore provide the selection information used to identify stars suitable for the variability search, while the 50\% completeness limits give a quantitative reference for the flare energies that can be reliably detected in each star. This information is useful beyond the variability analysis presented in this work. For any star in the sample, the injection-recovery results define the energy regime where a flare search is sensitive. Future studies interested in individual targets can therefore use these completeness limits to assess whether the absence of detected flares at a given energy is physically meaningful, or simply reflects the detection threshold of the TESS data for that star.

Also included in Table~\ref{tab:full_sample} are the mean and median scores from our visual search of the subset of 1775 stars which underwent the visual inspection procedure described in Section~\ref{sec:var_finding}. 

\begin{deluxetable*}{cccccccc}[ht]
\tablecaption{Properties for 14k stars searched in our CVZ sample for flares \label{tab:full_sample}}
\tablehead{
\colhead{TIC ID} &
\colhead{$N_{\rm sectors}$} &
\colhead{Baseline} &
\colhead{50\% Completeness Limit} &
\colhead{$N_{\rm flares}$} &
\colhead{$N_{\rm sectors}$ with $\geq 5$ flares} &
\colhead{Mean Score} &
\colhead{Median Score}
}
\startdata
364589441 & 35 & 1881.01 & 7.69   & 759  & 35 & 1.00  & 1  \\
364588501 & 35 & 1881.20 & 0.86   & 989  & 35 & 3.00  & 3  \\
350297825 & 35 & 1880.23 & 107.07 & 5340 & 35 & -0.80 & -1 \\
300969581 & 35 & 1881.03 & 5.29   & 637  & 35 & 0.33  & 0  \\
260127241 & 35 & 1880.20 & 6.67   & 596  & 35 & 1.67  & 2  \\
\enddata
\tablecomments{Baseline is given in days, and the 50\% completeness limit is given in seconds. The score columns summarize the visual classifications assigned during inspection. This table is shown in part while the full table will be made available electronically.}
\end{deluxetable*}

\begin{deluxetable*}{ccccccccc}[ht]
\tablecaption{Vetted sample Results}
\label{tab:Gold_Samp}
\tablehead{
\colhead{TIC ID} &
\colhead{Classification} &
\colhead{$m_T$\tablenotemark{\textdagger}} &
\colhead{$T_{\rm eff}$\tablenotemark{\textdagger}} &
\colhead{$P_{\rm rot}$} &
\colhead{$\langle$FFD Slope$\rangle$} &
\colhead{IQR($\beta$)} &
\colhead{IQR($N_{\rm flares}$)} &
\colhead{IQR($log(F_{\rm flare}/F_{\rm bol})$)}
}
\startdata
167344043 & Best      & 8.6163  & 5597 & 2.909 & 1.025 & 0.366 & 19.5 & 0.473 \\
364588501 & Best      & 9.2016  & 5605 & 2.279 & 0.825 & 0.279 & 17.0 & 0.495 \\
38827910  & Best      & 9.7510  & 5304 & 3.538 & 0.958 & 0.415 & 8.0  & 0.642 \\
25132999  & Best      & 10.8124 & 3945 & 5.172 & 1.072 & 0.282 & 11.0 & 0.464 \\
350520348 & Candidate & 8.4316  & 5032 & 1.878 & 1.214 & 0.230 & 8.5  & 0.528
\enddata
\tablenotetext{ $\textdagger$ }{
TESS apparent magnitude ($m_T$) and stellar effective temperature $T_{\rm eff}$ are taken from \citet{stassun_revised_2019}.
}
\tablecomments{$P_{\rm rot}$ is given in days. The IQR columns give the interquartile range of each metric across TESS sectors in Cycles 1-7. This table is shown in part, while the full table will be made available electronically. }
\end{deluxetable*}

Next, we present the subset of stars that received the strongest visual scores in the Cycle 1–5 analysis. Specifically, we define a Vetted sample of 75 stars with both a mean and median visual score greater than 2.0 in Table~\ref{tab:full_sample}. For these stars, we performed the additional processing described in Section~\ref{sec:cycle7_add_on_and_vis}, incorporating all available TESS data through Cycle 7, and repeated the visual inspection on the updated diagnostic products. The results of this final classification step are summarized in Table~\ref{tab:Gold_Samp}.

In addition to the final visual classification, Table~\ref{tab:Gold_Samp} reports several basic stellar and flare-variability properties for each candidate. We include the TESS magnitude and effective temperature from the TIC \citep{stassun_revised_2019}, along with our estimated photometric rotation period. We note that this period is measured from the dominant peak in the Lomb–Scargle periodogram of the GP-modeled light curve, and should therefore be interpreted as an approximate photometric modulation timescale rather than a definitive rotation period in every case. Additionally, in practice, our period sensitivity is limited by the $\sim$27 day duration of individual TESS sectors and by the requirement that the modulation be coherent enough to be recovered within the available baseline. Consequently, each of the candidates reported here have estimated periods shorter than $\sim$10 days. Although for vetted systems, these estimates seem to be consistent with those in the literature for these stars.

In Table~\ref{tab:Gold_Samp}, we also report the full-sample FFD slope and several summary measures of the time variability in the flare metrics. In particular, we list the interquartile ranges of $\beta$, flare count, and integrated flare power, across the analyzed sectors. These quantities are not used as a basis for classification, but provide a compact summary of how strongly each metric varies over time.

From the Vetted sample, we identify four stars in the ``Best'' category and 13 stars in the ``Candidate’’ category. The ``Best'' systems show the clearest evidence for long-term changes in flare behavior, with variability that is coherent across multiple metrics and resembles the behavior expected from an activity cycle. The ``Candidate’’ systems also show evidence for long-term flare variability, but either lack clear cyclical structure or retain some ambiguity due to sampling, uncertainties, or light-curve quality. All of the 17 stars in the top two categories span FFD slopes from approximately 0.8 to 1.55, while slopes greater than 2 were good indicators of ``Sneaky Trash''. 

We acknowledge that the ``Best'' classification is subjective, since it reflects the systems that stood out most clearly during visual inspection rather than the output of a quantitative metric. This subjectivity is also part of its value. The strongest candidates are the stars where multiple diagnostics tell the same story by eye, with coherent changes in the FFDs, $\beta$, flare counts, and integrated flare power. These systems therefore provide a practical benchmark for what convincing long-term flare variability looks like in the TESS data, even as the larger ``Candidate’’ category captures cases where the evidence is less unambiguous. For the rest of this work, we refer to this sample of 17 ``Candidate'' and ``Best'' stars as our candidates. 

For each of our candidate stars, we include the 6-panel diagnostic figure as a Figure Set in Appendix~\ref{appendix_1}. 
We now turn to further characterize our 17 candidates in more detail below.



\section{Flare Activity Variation} \label{sec:best_discussion}

The 17 stars in our candidate sample all show evidence for long-term changes in flare activity, but they do not all show the same temporal behavior. Some systems exhibit smooth rises or declines over the TESS baseline, while others show peaks, inflection points, or more irregular multi-epoch structure. In this section, we first discuss TIC 167344043, which provides the clearest example of solar-like flare-cycle behavior in our sample. We then consider the broader candidate population, focusing on how the amplitude and morphology of the flare variability change across stellar temperature, rotation period, and Rossby number.

\subsection{Examples of Solar-Like Behavior} \label{sec:poster_child}

\begin{figure*}[!ht]
    \centering
    \includegraphics[width=0.95\textwidth]{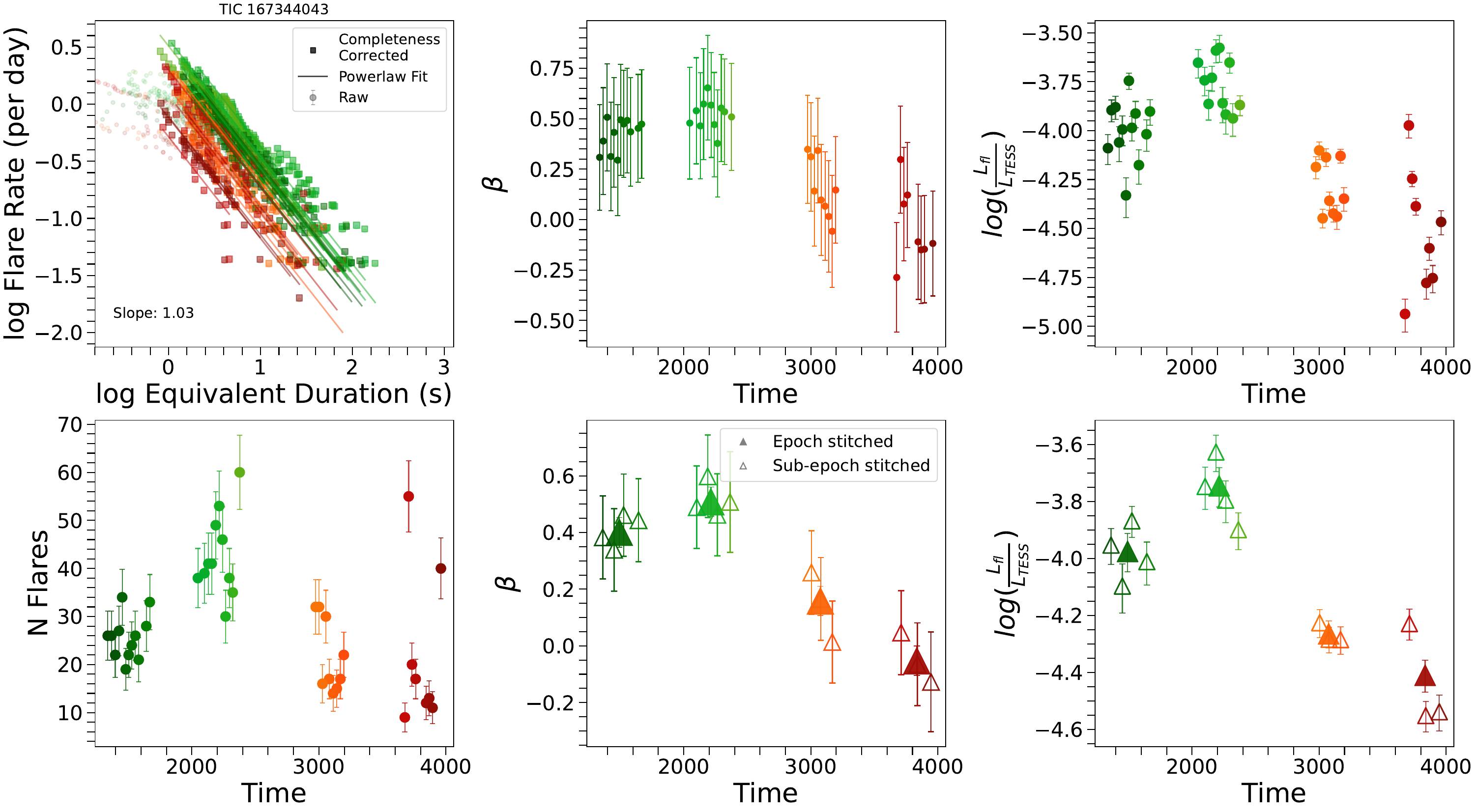}
    \caption{Diagnostic figure (same as Figure~\ref{fig:good_diagnostic_plot}) for TIC 167344043, a G4 type, super-flare star, with a rotation period of $\sim2.9$ days. In each panel, we see a a clear peak in the flare activity around ${\rm BTJD}\sim2200$, followed by a sustained decline, which suggests at least a 9 year cycle period. This activity cycle is potentially analogous to the suns, despite the much faster rotation and higher activity level.}
    \label{fig:poster_child}
\end{figure*}

Of the stars in our ``Best’’ category, TIC 167344043 provides the clearest example of the behavior we set out to search for. TIC 167344043 is a G4 dwarf with $T_{\rm eff}\sim5600$ K \citep{stassun_revised_2019}, similar in effective temperature to the Sun, but with a much shorter estimated rotation period of $\sim2.9$ days \citep[e.g.,][]{gunther_stellar_2020}. Even with this rapid rotation, the long-term evolution of its flare activity shows a strikingly solar-like, long duration, pattern. 

The 6-panel diagnostic plot for TIC 167344043 is shown in Figure~\ref{fig:poster_child}. In each panel, we see a rise in flare activity peaking around ${\rm BTJD}\sim2200$, followed by a sustained decline. During the high-activity epoch, shown in green, the FFD in the upper-left panel lies systematically above the later orange and red epochs, indicating a higher flare rate at fixed equivalent duration. Since each sector is fit with a fixed-slope power law, this vertical shift corresponds to the elevated values of $\beta$ seen near the activity maximum. The same structure appears across the other diagnostics. $\beta$, flare count, and the integrated flare power all peak near ${\rm BTJD}\sim2200$, followed by a long decline through the remainder of the TESS baseline. The stitched-sector measurements follow the same trend, showing that the behavior persists when neighboring sectors are combined.  

This behavior closely resembles the way flares follow the solar magnetic cycle, where eruptive activity is concentrated near activity maximum and becomes less frequent during the declining phase. While in this work we refrain from making uncertain cycle period estimations (See Section~\ref{sec:no_preferred_morphology}), for TIC 167344043, sinusoidal fits to the time evolution of the flare diagnostics favor periods of at minimum $\sim9$ yr. However, the current TESS baseline captures only the activity maximum and subsequent decline, so this period should be interpreted as a lower limit rather than a bona fide cycle measurement. Additional TESS observations will be needed to determine if and when the flare activity begins to rise again, as expected for a repeating magnetic cycle.

The flare behavior observed from TIC 167344043 is also broadly consistent with the range of activity-cycle phenomenology seen in long-baseline Mount Wilson Ca II H$\&$K monitoring \citep{duncan_ca_1991,baliunas_chromospheric_1995,hall_activity_2007}, where stars with Solar-levels of chromospheric emission show cyclic, monotonic, or irregular changes in magnetic activity. However, TIC 167344043 extends this type of decade-scale magnetic variability into a more active regime than the present-day Sun. Specifically, while TIC 167344043 has a similar effective temperature to the Sun, its much shorter rotation period suggests a younger star with a more magnetically active or ``saturated'' dynamo \citep[e.g.,][]{skumanich_time_1972,barnes_rotational_2003}. 

Further, TIC 167344043's identification as a superflare star, with flares reaching energies of order $10^{34}$ erg \citep{doyle_superflares_2020}, implies that at least some of its active regions are capable of storing and releasing substantially more magnetic energy than typical solar active regions. Producing such energetic events likely requires larger or more complex magnetic structures, such as large starspots or extended active-region complexes \citep{maehara_superflares_2012,notsu_superflares_2013,shibayama_superflares_2013,maehara_statistical_2015, tu_superflares_2020, tu_superflares_2021}. The presence of a solar-like rise and decline in TIC 167344043 is therefore particularly interesting because the long-term modulation appears to persist in a much different dynamo regime.

\begin{figure*}[ht]
    \centering
    \includegraphics[width=0.95\textwidth]{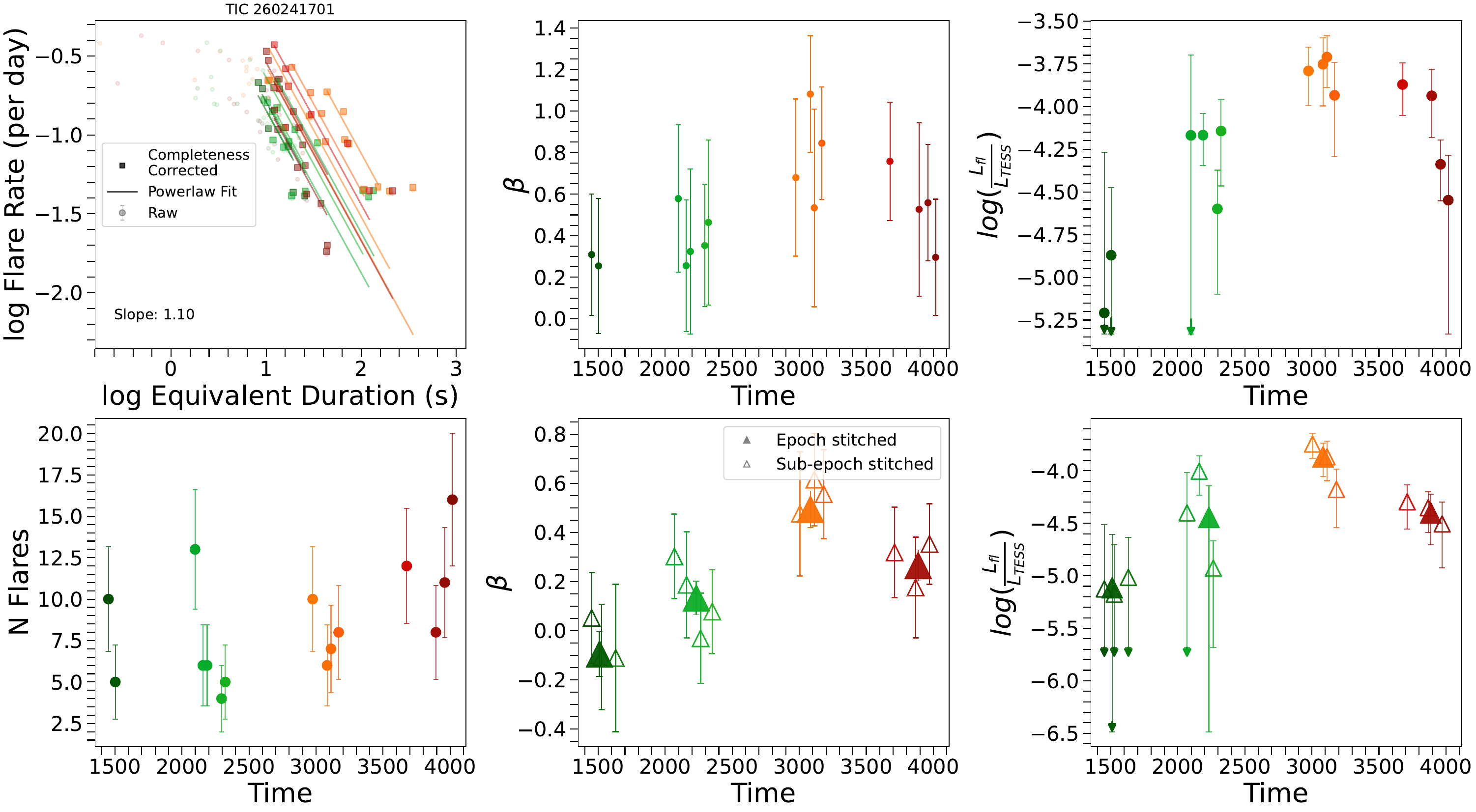}
    \caption{Diagnostic figure (same as Figure~\ref{fig:good_diagnostic_plot}, and Figure~\ref{fig:poster_child}) for TIC 260241701. In the each panel except number of flares, we see a peak in the activity around ${\rm BTJD}\sim3100$. This activity cycle is potentially analogous to the Sun, despite the cooler effective temperature and faster rotation rate.}
    \label{fig:second_example_less_clear}
\end{figure*}

While the majority of our ``Best'' candidates naturally also show high specific flare rates, we do see evidence for long term variation in less active stars. For example, TIC 260241701 ($T_{eff}\sim 3500$ K,  $P_{rot}\sim6$ days) exhibits an average of fewer than 10 flares per Sector, yet shows similar activity variation as TIC 167344043 above. 
During the third epoch around 3100 days, there is an inflection point in the flare metrics, except number of flares, as seen in Figure~\ref{fig:second_example_less_clear}. Therefore, as is clear in the FFD, during this epoch, there were more energetic flares, opposed to an increase in flare rate. Additional epochs of data will further help determine if there is cyclical behavior, or if we happened to catch a momentary stochastic increase in flare activity. If cyclical, TIC 260241701 would be another particularly interesting case of a rapidly rotating, lower mass star with a decade long activity cycle, and with a specific flare rate closer to what we observe for the Sun \citep[e.g.][]{shibayama_superflares_2013}.

\subsection{A Range of Flare-Variability Morphologies} \label{sec:no_preferred_morphology}

\begin{figure*}[ht]
    \centering
    \includegraphics[width=0.95\textwidth]{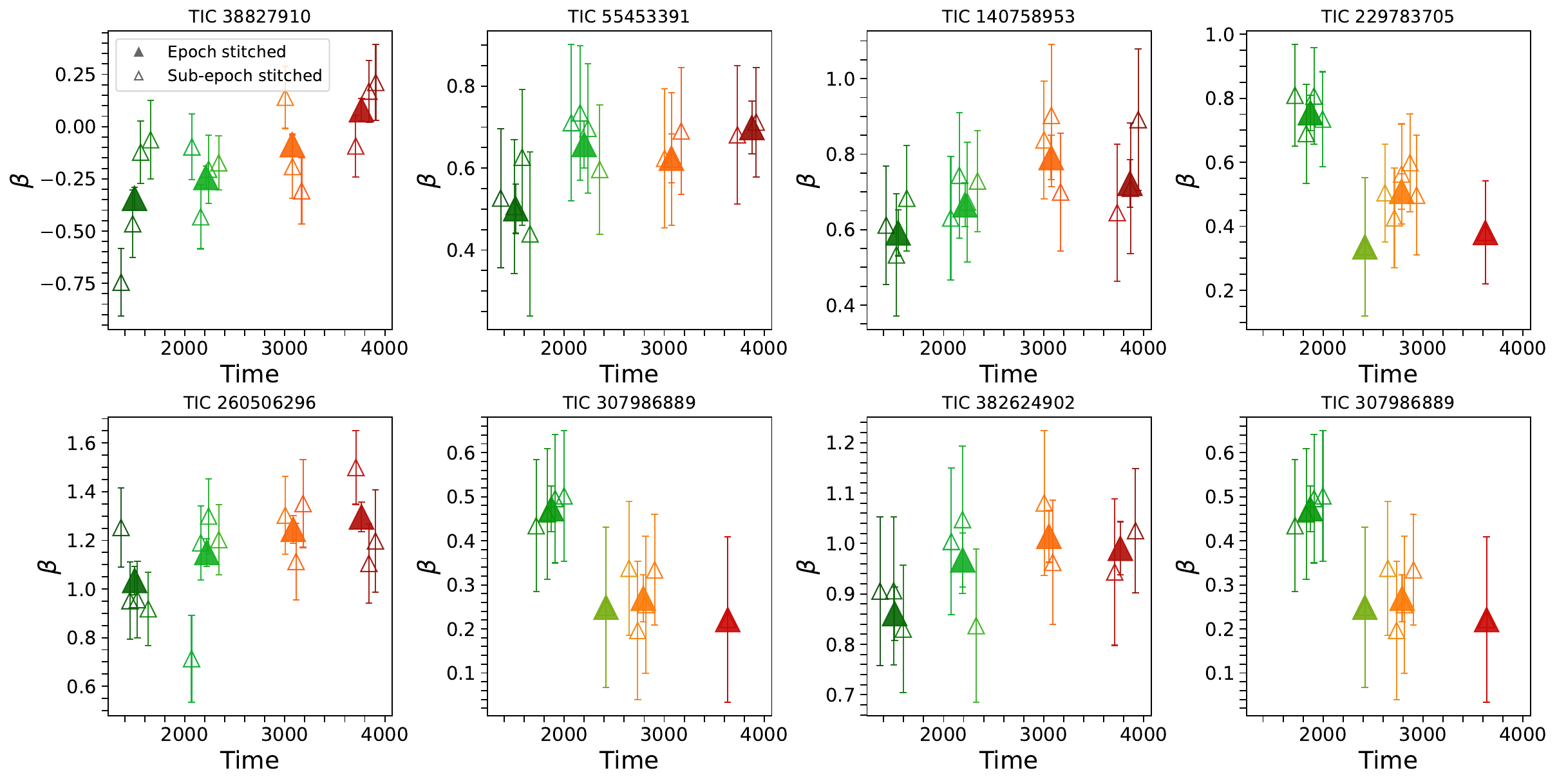}
    \caption{Binned $\beta$ evolution for eight candidates with relatively smooth or monotonic long-term behavior. Each panel shows the evolution of the FFD normalization $\beta$ as a function of time, using both epoch-stitched and sub-epoch-stitched measurements. This behavior could be consistent with activity cycle periods much longer than the TESS baseline probed here.}
    \label{fig:monotonic_beta}
\end{figure*}

Long-baseline chromospheric monitoring has long shown that the variability of stellar surface magnetic fields is diverse, with stars exhibiting cyclic, irregular, flat, and trend-like behavior, rather than a single activity morphology \citep{baliunas_chromospheric_1995,hall_stellar_2008, isaacson_california_2024}. The time-domain behavior of our 17 candidates is similarly diverse. In Figure~\ref{fig:monotonic_beta}, we show the binned $\beta$ evolution for eight stars whose long-term behavior is relatively monotonic or smooth variation. This behavior could be consistent with activity cycle periods much longer than the TESS baseline probed here. Some systems show a gradual increase in $\beta$ over the TESS baseline, while others show a decline from an earlier high-activity state. These trends are clearest in the stitched measurements, which reduce the sector-to-sector scatter from stochastic flare sampling while preserving the broad evolution in the FFD normalization.  

\begin{figure*}
    \centering
    \includegraphics[width=0.95\textwidth]{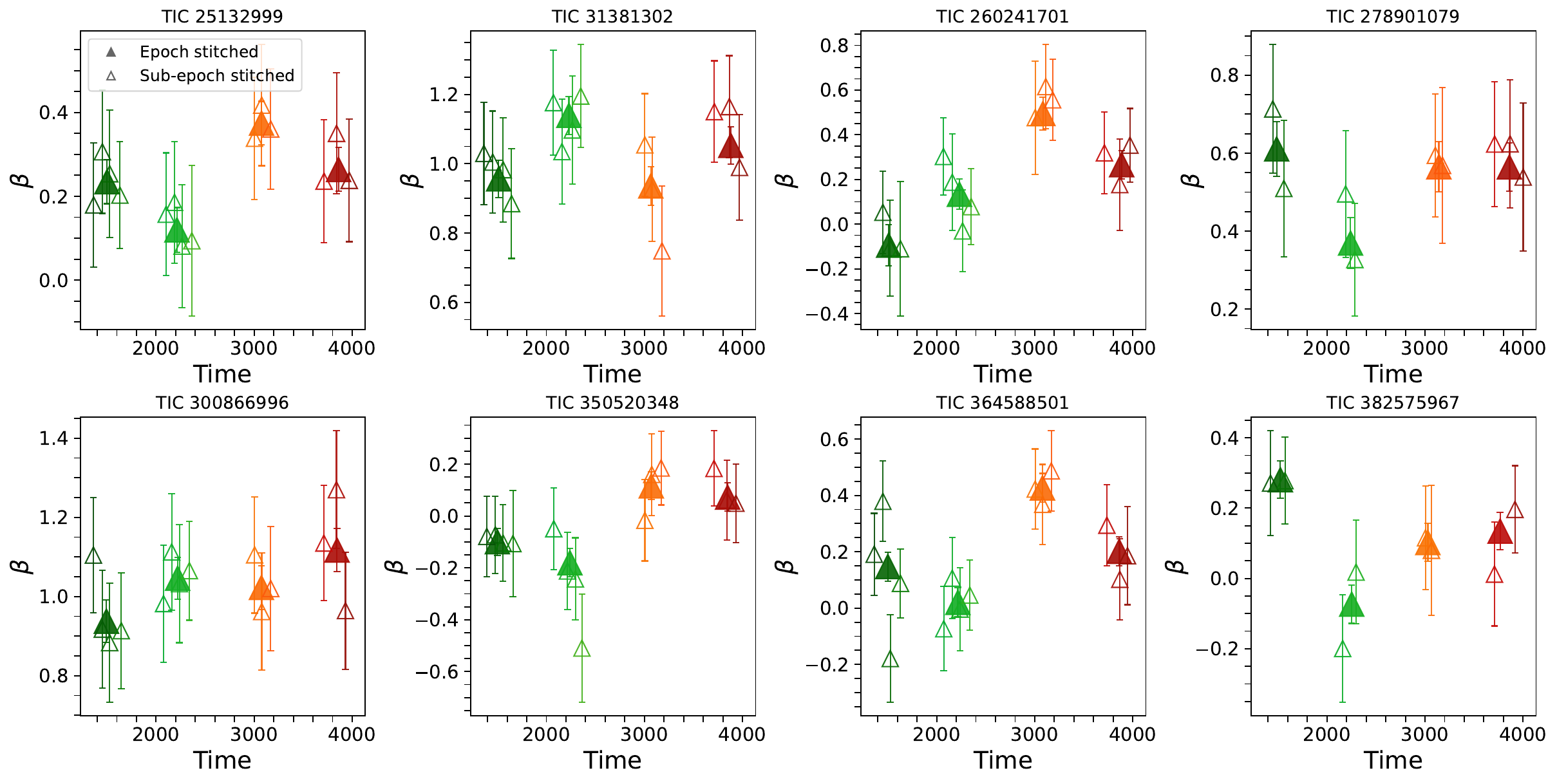}
    \caption{Binned $\beta$ evolution for eight candidates with more structured long-term behavior. As in Figure~\ref{fig:monotonic_beta}, each panel shows epoch-stitched and sub-epoch-stitched measurements of the FFD normalization. These stars show apparent inflection points, intermediate minima/maxima, or changes in the direction of the long-term trend.}
    \label{fig:inflection_beta}
\end{figure*}

In contrast, Figure~\ref{fig:inflection_beta} shows eight candidates with more structured behavior, including stars with apparent inflection points, intermediate maxima, or changes in the direction of the long-term trend. Some stars show behavior that could plausibly represent part of either a short, or long timescale activity cycle. Others may reflect shorter-lived changes in active-region coverage, stochastic clustering of large flares, or incomplete sampling of a more complex magnetic cycle.

Within the subset of candidates with potentially short activity-cycle periods (i.e., $\sim2-6$ years) we do not find an obvious trend with either rotation period or effective temperature. For example, TIC 31381302 is the fastest rotator in our sample, with an estimated rotation period of 0.165 days (consistent with the measurement from \citet{murray_study_2022}) and has an effective temperature of only $\sim2800$ K. However, its temporal flare-rate behavior is broadly similar to that of TIC 364588501, a much warmer star with $T_{\rm eff}\sim5600$ K and an order of magnitude longer rotation period of $P_{\rm rot}\sim2.8$ days. 

While previous activity-cycle studies provide useful context, they are not a direct expectation for our candidates. Some studies have reported correlations between cycle properties and stellar parameters such as rotation period or age \citep{bellotti_bcool_2025,metcalfe_stellar_2026}, while long-baseline chromospheric monitoring has simultaneously shown that activity cycles can be cyclic, monotonic, irregular, or difficult to classify \citep[e.g.,][]{wilson_probable_1963,wilson_chromospheric_1978,baliunas_chromospheric_1995}. Our flare-based sample shows a similar diversity of long-term behavior, but probes a much more rapidly rotating and flare-active regime than the canonical stellar-activity-cycle targets. These 17 stars are therefore not expected to correspond directly to either the present-day Sun or the older, more slowly rotating stars that dominate slowly rotating chromospheric samples \citep[e.g.,][]{isik_scaling_2023, isaacson_california_2025}.

For our sample, we do not see two full phases for any activity cycle, which is required for period determinations. We therefore refrain from making claims about activity cycle period in this work, or explicitly commenting on potential activity cycle, rotation period relations. However, as more TESS data becomes available, and full cycles phases able to be probed, these types of analysis will become more possible.

Instead, the value of this sample is the identification of candidate magnetic cycles in the active, rapidly rotating regime where stellar dynamos are still evolving. If these long-term flare variations are confirmed as activity cycles, they will provide observational constraints on when cycle-like behavior first emerges, how they depend on rotation and activity level, and help to determine whether the Sun is best viewed as a typical middle-aged example or as one point in a broader continuum of magnetic-cycle behavior.

\subsection{Variety of Stellar Properties} \label{sec:no_preferred_stellar_property}

\begin{figure*}
    \centering
    \includegraphics[width=0.90\textwidth]{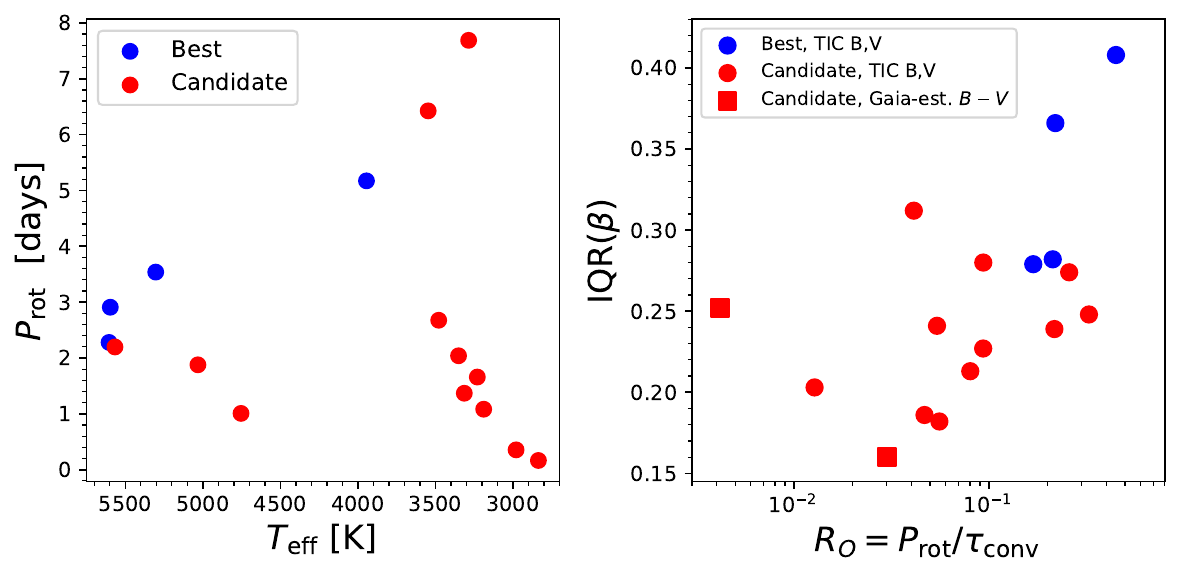}
    \caption{Stellar properties and flare-variability amplitudes for the final candidate sample. Left: effective temperature versus estimated rotation period for the Best'' and Candidate’’ stars. Right: Rossby number, $Ro=P_{\rm rot}/\tau_{\rm conv}$, versus IQR($\beta$), the interquatile range of sector $\beta$ values over the TESS baseline. Rossby numbers are computed using the \citet{noyes_rotation_1984} $B-V$-dependent convective turnover time prescription. Circles indicate stars with $B-V$ computed from TIC $B$ and $V$ magnitudes \citep{stassun_tess_2018}, while squares indicate stars for which $B-V$ was estimated from Gaia $G_{\rm BP}-G_{\rm RP}$ \citep{collaboration_gaia_2021}.}
    \label{fig:rossby_beta_iqr}
\end{figure*}

While Section~\ref{sec:poster_child} focused on variations that appeared to be particularly solar-like (i.e. long-term or decades timescale variations) for G dwarfs, our sample finds such variations across a range of spectral types. However, we do not see a common temporal morphology clearly emerging as a function of stellar properties. Figure~\ref{fig:rossby_beta_iqr} summarizes the basic parameter space for each the 17 candidate stars. The left panel shows the effective temperature\footnote{We note that TIC 382624902 doesn't have an effective temperature from \citet{stassun_revised_2019}, and therefore isn't plotted.} and estimated rotation period, colored by our ``Best'' and ``Candidate'' classifications. The sample includes both rapidly rotating cool M-type stars, and warmer G-type stars, indicating that the variability identified by our search is not confined to a single spectral type. 

The coolest stars in our sample are particularly interesting, as we have six stars have $T_{\rm eff}<3400$ K, including two with $T_{\rm eff}<3000$ K. Since the fully convective boundary occurs near $T_{\rm eff}\sim3200-3400$ K \citep{chabrier_structure_1997,baraffe_closer_2018,rabus_discontinuity_2019}, these stars may probe flare variability at, or below, a major transition in stellar interior structure. 

In solar-type dynamos, the tachocline at the interface between the radiative interior and convective envelope has long been considered an important shear layer for organizing large-scale magnetic-field generation \citep{spiegel_solar_1992,charbonneau_dynamo_2010}. Fully convective stars lack such an interface layer, yet observations and simulations show that these turbulent dynamos can still sustain strong and globally organized magnetic fields, and have rotation-dependent magnetic activity levels \citep{browning_simulations_2008,wright_solar-type_2016,wright_stellar_2018}. Long-term flare variability near the fully convective boundary therefore offers a direct way to test whether cycle-like magnetic behavior persists across this major transition in stellar dynamo structure. If such activity cycles for these candidates are confirmed, they provide evidence that long-term magnetic cycles can be maintained by a distributed convective dynamo operating throughout the stellar interior, even without a tachocline.

We further characterize the stars in our sample by their Rossby number. We estimate the Rossby number using the classic convective turnover time prescription of \citet{noyes_rotation_1984} such that,
\begin{equation}
    Ro = \frac{P_{\rm rot}}{\tau_{\rm conv}},
\end{equation}
where $P_{\rm rot}$ is the estimated photometric rotation period ,and $\tau_{\rm conv}$ is the convective turnover time. We compute $\tau_{\rm conv}$ using the color-dependent prescription from \citet{noyes_rotation_1984}. Specifically, defining
\begin{equation}
    x = 1 - (B-V),
\end{equation}
the convective turnover time in days is given by
\begin{equation}
    \log_{10}(\tau_{\rm conv}) =
    \begin{cases}
    1.362 - 0.166x + 0.025x^2 - 5.323x^3, & x > 0, \\
    1.362 - 0.14x, & x < 0.
    \end{cases}
\end{equation}

The right panel of Figure~\ref{fig:rossby_beta_iqr} compares the Rossby number, $Ro$, to the interquartile range (IQR) of $\beta$, which we use as a compact measure of the amplitude of the time-variable flare information.  

For 15 of the 17 stars in our sample, $B-V$ color is computed by the $B$ and $V$ apparent magnitudes available in the TIC \citep{stassun_tess_2018}. The remaining two stars lack TIC $B$ and $V$ values, but have Gaia $G_{\rm BP}$ and $G_{\rm RP}$ magnitudes. For these stars, we estimate an approximate the $B-V$ color using Gaia-to-Johnson photometric transformations from the literature \citep{jordi_gaia_2010,evans_gaia_2018,pancino_gaia_2022}. Because the \citet{noyes_rotation_1984} relation is calibrated in Johnson $B-V$, these two Gaia-based Rossby numbers are marked as squares in Figure~\ref{fig:rossby_beta_iqr}. As these Rossby numbers depend on heterogeneously derived colors and approximate photometric rotation periods, we treat them as qualitative diagnostics rather than precise stellar parameters.

While IQR($\beta$) has not been calibrated directly to the chromospheric or coronal activity indicators used in classical activity–Rossby studies (such as $R'_{\rm HK}$ or $R_X$ \citealt{wright_stellar-activity-rotation_2011}), IQR($\beta$) quantifies the range of flare activity observed in the FFD across the TESS baseline. This comparison is a qualitative diagnostic of how flare-variability amplitude relates to rotation and convection without assuming an activity cycle period, rather than as a direct activity–Rossby calibration.

Within this small sample, we find a modest trend between Rossby number and the amplitude of the flare-activity variation. Stars with the smallest Rossby numbers tend to show relatively modest values of IQR($\beta$), while the largest IQR($\beta$) values occur at larger Rossby numbers. While this trend should not be over-interpreted, it is suggestive that the strongest year-to-year changes in flare activity do not occur among the most rapidly rotating, lowest-Rossby-number stars. A larger sample is needed to determine whether this behavior reflects a real dynamo trend or simply the selection effects of our flare-detection and visual-classification procedure.

This apparent lack of large $\beta$ variations among the lowest-Rossby-number stars could reflect astrophysical differences in how flare activity varies in rapidly rotating stars that naturally reflects the broader context of the rotation–activity literature. While rapidly rotating stars are generally more magnetically active \citep{wright_stellar-activity-rotation_2011, fritzewski_detailed_2021}, and have higher flare activity \cite{davenport2019}, 
X-ray studies of low-mass stars show that fast rotators occupy a saturated regime whose behavior can depend on stellar mass and sample selection \citep[e.g.,][]{magaudda_relation_2020,magaudda_first_2022}. Since IQR($\beta$) measures the amplitude of the observed flare-activity variation rather than the absolute activity level, a star can be highly active on average while showing comparatively modest modulation in that activity. 

\begin{figure}[t]
    \centering
    \includegraphics[width=0.475\textwidth]{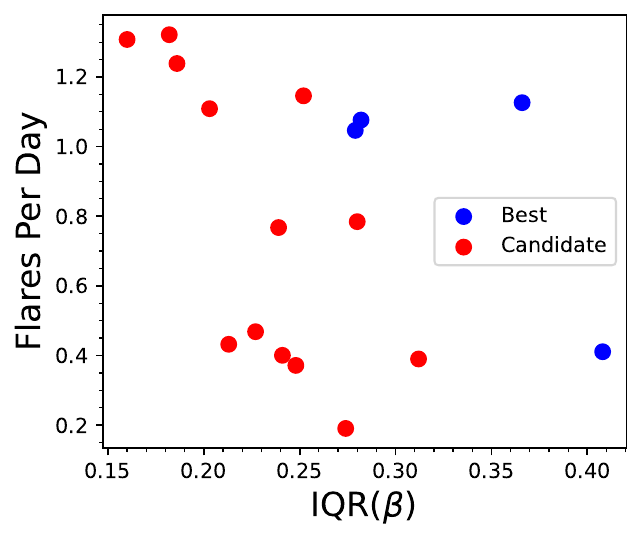}
    \caption{The flares per day for each candidate, compared to the IQR($\beta$), where our ``Best'' stars are shown in blue, like in Figure~\ref{fig:rossby_beta_iqr}. Of our candidates, stars with the most flares per day seem to have the smallest variation. }
    \label{fig:beta_iqr_vs_flares_per_day}
\end{figure}

This relation is broadly recovered in Figure~\ref{fig:beta_iqr_vs_flares_per_day}, where we plot each stars' flares per day across all available sectors (i.e. the mean flare rate), compared against the IQR($\beta$). While there is not a strong trend, we note the two stars with the highest mean flare rate have the smallest variation (i.e. IQR($\beta$)), while the star with the largest IQR($\beta$), has the fewest flares per day. Although the present sample of 17 stars is too small to establish whether this apparent structure represents a population-level trend, this comparison highlights a useful parameter space for future flare-cycle searches, where a larger sample might be able test how the amplitude of long-term flare variability depends on the absolute flare activity level.


\section{Discussion} \label{sec:discussion_section}

\subsection{Comparison to Previous Studies}\label{sec:comparison}

Studies that search for time-variable flare activity in large stellar samples are still relatively rare. Most flare surveys to date focus on average flare rates, flare energy distributions, or population-level trends with stellar properties, rather than asking whether individual stars exhibit significant changes in flare activity over time. Among the set of papers that address this question, methodology differences arise from the current work in the adopted time binning, activity metrics analyzed, and treatment of uncertainties. 

\citet{davenport_10_2020} provide useful context by showing both the growing interest in flare-based cycle searches and the fact that some highly active stars show little evidence for long-term flare-rate evolution. \citet{scoggins_using_2019} presented evidence for potential long term variation between Kepler and TESS which was later shown to be a bi-product of cadence and bandpass differences in the two telescopes \citepalias{wainer_searching_2024}.

\citet{feinstein_evolution_2024} is the most direct comparison for our work. They used five years of TESS observations and searched for annual flare-rate variability in young GKM stars, identifying 11 candidate stars, 8 of which were in our CVZ sample. For each of these stars, we added Cycle 7 data as in our Vetted sample (Section~\ref{sec:cycle7_add_on_and_vis}) for best comparison. 
While each of these 8 stars made it to visual inspection, none appear in our final Vetted sample. However, this does not mean that the two analyses are in direct disagreement. When we generate a like-for-like comparison using the same coarse year-long binning of only integrated flare power, we recover the broad activity enhancements that formed the basis of their classifications. In that sense, our results are consistent with \citet{feinstein_evolution_2024} at the level of the binned trends they identified. 

The difference emerges once the data are examined in more than one metric, with a more explicit treatment of the uncertainties. Our analysis incorporates the full sector-level information, includes propagated uncertainties on each flare-activity metric, and evaluates the behavior of multiple diagnostics together rather than relying on a single measure of flare activity. Under this stricter treatment, the apparent year-to-year changes in these overlapping stars generally become less compelling. In particular, for most cases, once the sector-level scatter, the metric uncertainties, and the behavior of $\beta$, the integrated flare power, and the flare counts are considered together, the evidence supporting classifying these systems as robust variability candidates becomes much less clear.

One notable exception is TIC 260351540. In this case, the additional TESS data obtained after the time span considered by \citet{feinstein_evolution_2024}\footnote{\citet{feinstein_evolution_2024} used data from Sectors 1 through 67.} reveals an apparent burst of low-energy flares. The decline observed in \citet{feinstein_evolution_2024} is still clear in our data, but the additional sectors remove the long term trend. We analyze this short term enhancement in detail in an upcoming work (Stuart et al. in preparation). 

While the flare analysis in \citet{feinstein_evolution_2024} provides the most direct comparison to our results, flare rates are only one observable tracer of stellar magnetic variability. Star spots provide a complementary photometric tracer because changes in their coverage, contrast, and longitudinal distribution can alter the amplitude and coherence of rotational modulation. These spot-modulation signals are typically lower amplitude than flares, and require careful interpretation \citep{basri_information_2020}, but they can be measured from the same long-baseline space-based photometry used for flare searches. Since flares and spots both trace the evolving magnetic state of the stellar surface, the strongest activity-cycle candidates may plausibly appear in both flare-based and spot-based searches.

Long-baseline photometric searches for stellar activity cycles beyond the Sun are still relatively limited \citep{jeffers_stellar_2023}, but several studies have used changes in rotation-period signals or spot-modulation amplitudes to identify candidate cycles. For example, \citet{vida_looking_2014} searched for activity cycles in late-type Kepler stars using time-frequency analysis, and \citet{chahal_photometric_2025} combined Kepler and TESS-era photometry to search for photometric activity cycles in rapidly rotating stars. Their samples do not overlap with ours, so they do not provide a direct object-by-object comparison, but they demonstrate that long-term changes in spot modulation offer an important complementary view of magnetic variability in the same broad class of active stars.

One recent study however, \citet{ramsay_searching_2024}, searched for activity cycles using TESS photometry of low-mass stars in the CVZ. While their analysis is not a like-for-like comparison to ours, since it is based on rotational modulation rather than flare activity, it probes magnetic variability over a similar TESS baseline, and 25 of their 27 candidates were included in our visual inspection sample. This overlap provides a useful check on whether stars identified through spot-modulation variability also show evidence for time-variable flare activity in our analysis. 22 of the 25 \citet{ramsay_searching_2024} candidates we deemed to have no long-term trends in flare rate in our visual inspection. However, three stars are in our Vetted sample, two of which (TIC 353898013, and TIC 300866996) are classified as a ``Candidate'' while one was ``sneaky trash'' (TIC219773818). TIC 300866996 and TIC 353898013 are then the clearest case where the spot-modulation and flare-based diagnostics both point toward time-variable magnetic activity, and offer obvious systems for follow up study of magnetic activity. 


One final activity cycle study we compare against is that of \citet{isaacson_california_2024}, who performed long term chromospheric monitoring of calcium H and K lines to search for activity cycles. Of their 138 cycle candidates, two (TIC 232606278 and TIC 259237827) are in our full CZV sample, although neither had enough identified flares to qualify for visual inspection. Both stars are classified as K0 type stars, and are older, with activity cycle periods reported to be 12.5 and 5.5 years respectively \citep{isaacson_california_2025}. Therefore, the lack of identifiable flares in our study is not necessarily surprising, as the typical detection threshold for flares on brighter K0 stars may well be larger than flare energies produced by these stars, even if activity cycles are present, as we discuss further in Section~\ref{sec:why_not_more}. These results help support that the two methods of identifying activity cycle behavior (i.e., flare based, and chromospheric monitoring) are truly complimentary, probing different regimes of stellar type and properties. 

\subsection{Why Aren't There More Candidates? } \label{sec:why_not_more}





At first glance, identifying only a few dozen plausible flare-variability candidates from a sample of more than 14,000 stars may seem surprising. However, the relevant denominator is not the full number of stars observed by TESS, but the much smaller subset of stars for which this experiment is actually viable. Here we discuss some of the limitations to this viability, and discuss why 17 candidates is actually a reasonable number given the current technology.

Flare occurrence depends strongly on spectral type, age, rotation, and magnetic activity, and bandpass \citep[e.g.,][]{howard_characterizing_2023}. Even among M dwarfs, which are the most active and favorable stars for optical flare searches, only a few percent of field stars show detectable flares in a given survey \citep{davenport_kepler_2016, stelzer_flares_2022}. Thousands of stars with excellent sampling from the CVZ  will therefore likely only yield of order hundreds of bona fide flare stars to search for activity variations.

Even for stars that are flare active, detecting those flares remains challenging. Flare detection is particularly difficult for moderate to low flare rates, and with small amplitude (low energy) flares, exactly the regime where long-term stellar-cycle searches are most interesting. Automated tools such as \texttt{stella} make it possible to process thousands of light curves uniformly, but are far from perfect classifiers. As discussed, eclipsing binaries, pulsators, rapid rotators with sharp spot-modulation features, instrumental systematics, and low signal-to-noise light curves can all inhibit robust detection in any automated search. 

This flare finding hurdle is especially important for G and K dwarfs. These stars are the most natural analogs to the Sun and therefore among the most constraining targets for activity cycles. However, their white-light flares are harder to detect than those of active M dwarfs \citep[e.g.,][]{walkowicz_white-light_2011, davenport_kepler_2016}. G-dwarf candidates in our sample tend to be stars with unusually energetic flares, consistent with the broader superflare literature. Superflares on solar-type stars are typically defined as events with bolometric energies of order $10^{32}$–$10^{36}$ erg, and Kepler/TESS studies have shown that such events are rare but detectable on solar-type stars when the flare amplitudes are large enough \citep{maehara_statistical_2015,doyle_superflares_2020}. Stars with more solar-like flare energies may simply fall below the effective detection floor of our current TESS flare-finding pipeline, even if they do possess magnetic cycles.

This flare-detection threshold likely explains why two stars with chromospheric activity cycles identified by \citet{isaacson_california_2024} are not recovered in our search. In order to be in our sample, a star must produce flares that are frequent enough, energetic enough, and sufficiently well separated from the photometric noise to be measured over multiple sectors. It is therefore likely that some stars in our full CVZ sample host magnetic activity cycles (possibly detectable from chromospheric monitoring), but whose flares remain below our detection threshold. The 17 candidates identified here are the subset for which that variability rises above the noise, making them a realistic and intentionally conservative sample of stars with measurable long-term flare variability.

The opposite problem occurs for M dwarfs. They are numerous, magnetically active, and much easier targets for flare detection, but their magnetic behavior is not necessarily expected to be solar-like. Observations and models suggest that fully convective M dwarfs can host strong magnetic fields and may occupy multiple magnetic-topology states at similar masses and rotation rates, complicating simple expectations for activity cycles \citep{kitchatinov_magnetic_2014,ortiz-rodriguez_simulations_2023, kapyla_simulations_2023}. Thus, while M dwarfs provide the largest pool of detectable flares, it is not obvious that they should commonly show coherent, activity cycles. The presence of a few M dwarfs among our strongest candidates is therefore particularly interesting, raising the question of what distinguishes these stars from the many other active M dwarfs in the sample.  

A second selection effect arises from the time-domain morphology of the variability itself. Our initial visual search was performed using the first five TESS cycles, which provided roughly three observing epochs for most CVZ stars. With only three epochs, many systems showed ambiguous behavior, where it was very difficult to differentiate an apparent rise, decline, or intermediate maximum that could represent part of a coherent long-term trend, from stochastic flare-count fluctuations from a given epoch. The addition of Cycle 7 provided a valuable fourth epoch, which clarified the behavior of some systems, but also made others less compelling. 

The final sample therefore represents a specific realization of flare variability over the TESS epochs observed so far. Stars whose activity changed strongly during the first several epochs were easier to recognize, while stars that were relatively steady during those same epochs may still show interesting long-term behavior as the baseline grows. Poisson fluctuations in the number of detected flares per sector further amplify the morphology-selection effect, especially for stars near the threshold where only a small number of flares contribute to each FFD. Continued TESS monitoring will therefore be essential for distinguishing persistent magnetic-cycle-like evolution from stochastic epoch-to-epoch variability. As additional sectors become available, stars that were ambiguous or apparently quiet in the current baseline may emerge as strong candidates, increasing the sensitivity of this approach to a broader range of cycle phases and morphologies.

These combined effects make our final sample of 17 candidates both plausible and scientifically useful. These stars reflect the difficulty of detecting stellar activity cycles through flares, and a purposefully conservative and stringent target list for future study, rather than evidence that such cycles are intrinsically absent from the broader stellar population.

\section{Conclusion}
\label{sec:end}

In this work, we have searched over 14,000 stars in the TESS CVZ for stellar flares. We find 17 candidates that exhibit significant variability in their flare rate over a 7 year baseline. Our main conclusions are: 

\begin{itemize}
    \item We find evidence for a solar-like activity cycle operating on the G-dwarf TIC 167344043 based on our point-in-time flare activity measures. TIC 167344043 is fast rotating, with super-flare behavior, probing a new regime in stellar activity cycles.

    \item Within our 17 candidates, we find no preferred variability morphology. We find stars that exhibit monotonic behavior over the TESS baseline, possible evidence of an activity cycle with a period much longer than the TESS baseline, while also seeing stars with clear inflection points, with potential activity cycle periods of order $\sim5$ years. 

    \item Our 17 candidates include stars with a variety of effective temperatures and rotation periods, with no clear activity cycle morphology dependence on stellar properties. Within this sample are activity cycle candidates for M-dwarfs with effective temperatures that suggest they are fully convective, potentially providing constraints on dynamo models where an activity cycle can arise without the existence of a tachocline layer. We also find that stars which exhibit the largest variations in flare rate have the largest Rossby numbers, despite all stars being within the saturated regime. 

    \item Our 17 candidates are preferentially rapidly rotating, likely younger, and more active than the slowly rotating, solar-type stars that have anchored much of the classical activity-cycle literature. Long-term flare variability in these systems therefore provides a new way to test whether cycle-like magnetic behavior is already present while stellar dynamos are still evolving. These stars can therefore help constrain when organized, cycle-like activity first emerges, and how it depends on rotation and magnetic activity level.
\end{itemize}

\vspace{1cm}
Portions of this work were conducted while TW was a pre-doctoral fellow at the Center for Computational Astrophysics. TW thanks the Flatiron Institute and the Simons Foundation. The Flatiron Institute is funded by the Simons Foundation. Additionally, this work was greatly benefited from numerous conversations with Dax Felix, Lehman Garrison, Lionel Garcia.

This paper includes data collected by the TESS mission, which are publicly available from the Mikulski Archive for Space Telescopes (MAST).

We acknowledge support from the DiRAC Institute in the Department of Astronomy at the University of Washington. The DiRAC Institute is supported through generous gifts from the Charles and Lisa Simonyi Fund for Arts and Sciences, Janet and Lloyd Frink, and the Washington Research Foundation. 
G.T.M. acknowledges support from the National Science Foundation MPS-Ascend Post- doctoral Research Fellowship under grant No. 2402296.

This research was supported by the National Aeronautics and Space Administration (NASA) under grant number 80NSSC21K0362 from the TESS Cycle 3 Guest Investigator Program, grant number 80NSSC23K0155 from the TESS Cycle 5 Guest Investigator Program, grant number 80NSSC25K0115 from the TESS Cycle 7 Guest Investigator Program, and award number from the Astrophysics Data Analysis Program (ADAP).

This publication also uses data generated via the Zooniverse.org platform, development of which is funded by a generous grant from the Alfred P. Sloan Foundation and by custom grants from the contributing research teams.

\software{\texttt{astropy} \citep{astropy_collaboration_astropy_2013, astropy_collaboration_astropy_2018, astropy_collaboration_astropy_2022}, \texttt{Jupyter} \citep{kluyver2016jupyter}, \texttt{matplotlib} \citep{Hunter:2007}, \texttt{numpy} \citep{harris_array_2020}, \texttt{python} \citep{python}, \texttt{scipy} \citep{2020SciPy-NMeth, scipy_12522488}, \texttt{h5py} \citep{collette_python_hdf5_2014, h5py_7560547}, \texttt{stella} \citep{feinstein_stella_2020}, \texttt{tinygp} \citep{foreman-mackey_dfmtinygp_2024}, \texttt{jaxop} \citep{jax2018github},
\citet{openai2023gpt4}}

\appendix

\section{Figure Set and Data For 17 Candidates} \label{appendix_1}
For the 17 stars with long-term flare-rate variations indicative of candidate activity cycles, we provide the full set of six-panel diagnostic plots and the data products used to generate them. The online journal version includes these diagnostic plots as a figure set. The same figures, along with the plotting scripts and documentation, are available on GitHub at \href{https://github.com/tobin-wainer/Searching_For_Activity_Cycles_Using_Flares_Public_Interface/tree/main}{https://github.com/tobin-wainer/Searching\_For\_Activity\_Cycles\_Using\_Flares\_Public\_Interface/tree/main}. The underlying data products are available on Zenodo at \href{https://doi.org/10.5281/zenodo.21119254}{doi:10.5281/zenodo.21119254}. These data products are provided as HDF5 files and include the all flare-finding results, completeness data, FFD measurements, flare diagnostics for both sector level and stitched epochs, and all quantities used in the diagnostic figures. The GitHub repository includes a description of the HDF5 file structure, example code for accessing the data products, and scripts for reproducing the six-panel diagnostic plots.



\bibliographystyle{aasjournal}
\bibliography{tobins_references, software, additional}

\end{document}